\documentclass{ws-ijmpe}
\usepackage{tipa}
\begin{document}
\markboth{Jameel-Un Nabi}{$\beta$-decay of key titanium isotopes in
stellar environment}
\catchline{}{}{}{}{}
\title{$\beta$-decay of key titanium isotopes in stellar environment}
\author{\footnotesize Jameel-Un Nabi \footnote{Corresponding author.}}
\address{Faculty of Engineering Sciences, GIK Institute of Engineering
Sciences and Technology, Topi 23640, Khyber Pakhtunkhwa, Pakistan\\
jameel@giki.edu.pk}
\author{\footnotesize Irgaziev Bakhadir}
\address{Faculty of Engineering Sciences, GIK Institute of Engineering
Sciences and Technology, Topi 23640, Khyber Pakhtunkhwa, Pakistan\\
bakhadir@giki.edu.pk}
 \maketitle
\begin{history}
\received{(received date)}
\revised{(revised date)}
\end{history}
\begin{abstract}
Amongst iron regime nuclei, $\beta$-decay rates on titanium isotopes
are considered to be important during the late phases of evolution
of massive stars. The key $\beta$-decay isotopes during presupernova
evolution were searched from available literature and a microscopic
calculation of the decay rates were performed using the
proton-neutron quasiparticle random phase approximation (pn-QRPA)
theory. As per earlier simulation results electron capture and
$\beta$-decay on certain isotopes of titanium are considered to be
important for the presupernova evolution of massive stars. Earlier
the stellar electron capture rates and neutrino energy loss rates
due to relevant titanium isotopes were presented. In this paper we
finally present the $\beta$-decay rates of key titanium isotopes in
stellar environment. The results are also compared against previous
calculations. The pn-QRPA $\beta$-decay rates are bigger at high
stellar temperatures and smaller at high stellar densities compared
to the large scale shell model results.
This study can prove useful for the core-collapse simulators. \\ \\
PACS Number(s): 97.10.Cv, 26.50.+x, 23.40.-s, 23.40.Bw, 21.60.Jz
\end{abstract}

\section{Introduction}
Shortly after the discovery of the Gamow-Teller (GT) resonance,
Bethe and collaborators \cite{Bet79} suggested its importance for
stellar weak-interaction mediated reactions. The calculations of
stellar $\beta$-decays and electron captures are very sensitive to
the distribution and the total strength of the GT giant resonance.
The centroid of the GT distribution functions determines the
effective energy of the capture and decay reactions. Due to these
weak interaction processes the value of $Y_{e}$ (lepton-to-baryon
ratio) for a massive star changes from 1 (during hydrogen burning)
to roughly 0.5 (at the beginning of carbon burning)  and finally to
around 0.42 just before the collapse to a supernova explosion.
$\beta$-decay are relatively unimportant at the beginning of carbon
burning phase of massive stars ($Y_{e} \sim 0.5$). However it
becomes increasingly competitive for neutron-rich nuclei due to an
increase in phase space related to larger Q$_{\beta}$ values. In
earlier simulations of core-collapse studies $\beta$-decay of heavy
nuclei were not considered. However it was soon acknowledged that
during the early stages of the collapse, a strong $\beta$-decay rate
can contribute both to the cooling rate (via the stellar
antineutrinos produced) and to a much larger value of $Y_{e}$ at the
formation of homologous core. This in turn would result in a smaller
envelope and to a more energetic shock wave. Aufderheide and
collaborators \cite{Auf94} stressed on the importance of
$\beta$-decay rates in the iron core prior to the collapse.  The
authors also found in their calculation that the $\beta$-decay rates
were strong enough and at times larger than the competing electron
capture rates for $Y_{e} \sim 0.42 - 0.46$. The temporal variation
of $Y_{e}$ within the core of a massive star has a pivotal role to
play in the stellar evolution and a fine-tuning of this parameter at
various stages of presupernova evolution is the key to generate an
explosion. Collapse simulators world-wide find it challenging to
transform the collapse of the core of massive stars to an explosion
and to date there have been no successfully simulated spherically
symmetric explosions. Even the 2D simulations (addition of
convection) performed with a Boltzmann solver for the neutrino
transport fails to convert the collapse into an explosion
\cite{Bur03}. (It is worth mentioning that recently a few simulation
groups (e.g. Refs. \cite{Buras06,Bur06,Woo07}) have reported
successful explosions in 2D mode.) Additional energy sources (e.g.
magnetic fields and rotations) were also sought that might transport
energy to the mantle and lead to an explosion. World-wide
core-collapse simulators are still working hard to come up with a
convincing and decisive mode of producing explosions.

The calculation of stellar weak interaction rates has been performed
by a number of authors during 1960's and 1970's. The rates were
calculated using the best available physics of the time. However the
first breakthrough was achieved in 1980 when Fuller, Fowler and
Newman (commonly referred to as FFN) \cite{Ful80} used a simple
shell model to compute the location and strength of the GT
resonances for 226 nuclei in the mass range $21 \leq A \leq 60$. FFN
estimated the GT contributions to the rates by a parametrization
based on the independent particle model. Aufderheide et al.
\cite{Auf94} later extended the FFN work for heavier nuclei with A
$>$ 60 and took into consideration the quenching of the GT strength
neglected by FFN.

Experimentally it has been shown that for $(p,n)$ and $(n,p)$
reactions the 0$^{0}$ cross sections for such transitions are
proportional to the squares of the matrix elements for the GT
$\beta$ decay between the same states (e.g. Ref. \cite{Goo80}).
Results of such measurements disclosed that, in contrast to the
independent particle model, the total GT strength is quenched and
fragmented over many final states in the daughter nucleus caused by
the residual nucleon-nucleon correlations. Both these effects are
caused by the residual interaction among the valence nucleons and an
accurate description of these correlations is essential for a
reliable evaluation of the stellar weak interaction rates due to the
strong phase space energy dependence. The measured data from various
$(p,n)$ and $(n,p)$ experiments also revealed the misplacement of
the GT centroid adopted in the parameterizations of FFN and
subsequently used in the calculation of weak rates by Ref.
\cite{Auf94}. Since then theoretical efforts were concentrated on
the microscopic calculations of weak-interaction mediated rates of
iron-regime nuclide. Two such widely used models are the large-scale
shell model (LSSM)(e.g. Ref. \cite{Lan00}) and the proton-neutron
quasiparticle random phase approximation (pn-QRPA) theory  (e.g.
Ref. \cite{Nab04}). Both LSSM and pn-QRPA model later validated the
finding of Ref. \cite{Auf94}, on microscopic grounds rather than
phenomenological parametrization of GT centroids and strengths, that
$\beta$-decay rates are larger than electron capture rates for a
certain range of $Y_{e}$ values. The LSSM calculation, however, makes use of the so-called Brink's hypothesis in
the electron capture direction and back-resonances in the
$\beta$-decay direction to approximate the contributions from
high-lying excited state GT strength distributions. Brink's
hypothesis states that GT strength distribution on excited states is
\textit{identical} to that from ground state, shifted \textit{only}
by the excitation energy of the state. GT back resonances are the
states reached by the strong GT transitions in the inverse process
(electron capture) built on ground and excited states. $\beta$-decay rate calculation for
iron-regime nuclei are also important during the late stages of
stellar evolution.

Isotopes of titanium are amongst the key iron-regime nuclei that
play a key role in the developments during the late phases of
stellar evolution. A search was performed by authors of Ref.
\cite{Auf94} for most important electron capture and $\beta$-decay
nuclei after core silicon burning in massive stars. The authors
assumed that the temperatures within the iron core were high enough
for the strong and electromagnetic reactions to be in equilibrium.
The nuclear Saha equation was then used to compute the isotopic
abundances. They tabulated the 71 top $\beta$-decay nuclei averaged
throughout the stellar trajectory for $0.40 \leq Y_{e} \leq 0.5$
(see Table 26 therein). From this table, six isotopes of titanium,
namely $^{51,52,53,54,55,56}$Ti, were short-listed as those whose
$\beta$-decay rates were suggested to have a significant
contribution during and after core silicon burning phases in massive
stars. A microscopic calculation of stellar $\beta$-decay rates was
performed for these titanium isotopes using the pn-QRPA theory.
Because of the high temperatures prevailing during the presupernova
and supernova phase of a massive star, there is a reasonable
probability of occupation of parent excited states and the total
weak interaction rates have a finite contribution form these excited
states. The pn-QRPA theory allows a microscopic state-by-state
calculation of \textit{all} these partial rates and this feature of
the model greatly enhances the reliability of the calculated rates
in stellar matter. The pn-QRPA model can handle any arbitrarily
heavy system of nucleons as it has access to a luxurious model space
of up to $7\hbar\omega$ shells. The pn-QRPA model was successfully
used to calculate weak interaction rates on important iron-regime
nuclei (e.g. Refs. \cite{Nab04,Nab99,Nab05,Nab07a,Nab08,Nab09}).
Earlier Nabi and collaborators \cite{Nab07} presented a detailed
analysis of the calculation of stellar electron capture rates on
twenty two titanium isotopes. Later a calculation of neutrino and
antineutrino energy loss rates in massive stars due to isotopes of
titanium was also presented \cite{Nab10}. In this paper we present
the $\beta$-decay rates due to the above mentioned six isotopes of
titanium in stellar matter. This completes the analysis of weak
interaction rates of key titanium isotopes considered to be
important for the presupernova evolution of massive stars. These
microscopic rates can be of great utility for core-collapse
simulators. The next section discusses briefly the formalism and
presents the calculated $\beta$-decay rates. Comparison with
previous calculations is also presented in this section. Section 3
finally summarizes the main conclusions.

\section{Calculations and Results}
The Hamiltonian of the pn-QRPA model and its diagonalization were
discussed earlier in Ref. \cite{Nab07}. The $\beta$-decay rates of a
transition from the $i^{th}$ state of the parent to the $j^{th}$
state of the daughter nucleus are given by
\begin{equation}
\lambda ^{^{\beta} } _{ij} =\left[\frac{\ln 2}{D}
\right]\left[B(F)_{ij} + (g_{A}/g_{V})^{2} B(GT)_{ij}
\right]\left[f_{ij}^{\beta} (T,\rho ,E_{f} )\right]. \label{phase
space}
\end{equation}
The value of D was taken to be 6295s \cite{Yos88}. B(F) and B(GT)
are reduced transition probabilities of the Fermi and ~Gamow-Teller
(GT) transitions, respectively,
\begin{equation}
B(F)_{ij} = \frac{1}{2J_{i}+1} \mid<j \parallel \sum_{k}t_{\pm}^{k}
\parallel i> \mid ^{2}.
\end{equation}
\begin{equation}
B(GT)_{ij} = \frac{1}{2J_{i}+1} \mid <j \parallel
\sum_{k}t_{\pm}^{k}\vec{\sigma}^{k} \parallel i> \mid ^{2}.
\end{equation}
Here $\vec{\sigma}^{k}$ is the spin operator and $t_{\pm}^{k}$
stands for the isospin raising and lowering operator.

The $f_{ij}^{\beta}$ are the phase space integrals and are functions
of stellar temperature ($T$), density ($\rho$) and Fermi energy
($E_{f}$) of the electrons. They are explicitly given by
\begin{equation}
f_{ij}^{\beta} \, =\, \int _{1 }^{w_{m}}w\sqrt{w^{2} -1} (w_{m} \,
 -\, w)^{3} F(+ Z,w)(1- G_{-}) dw,
 \label{bd}
\end{equation}
In Eq. ~(\ref{bd}), $w$ is the total energy of the electron
including its rest mass. $w_{m}$ is the total $\beta$-decay energy,
\begin{equation}
w_{m} = m_{p}-m_{d}+E_{i}-E_{j},
\end{equation}
where $m_{p}$ and $E_{i}$ are mass and excitation energies of the
parent nucleus, and $m_{d}$ and $E_{j}$ of the daughter nucleus,
respectively. $F(+ Z,w)$ are the Fermi functions and were calculated
according to the procedure adopted by Gove and Martin \cite{Gov71}.
$G_{-}$ is the Fermi-Dirac distribution function for electrons.
Details of the calculation of reduced transition probabilities can
be found in Ref. \cite{Nab04}. Construction of parent and daughter
excited states and calculation of transition amplitudes between
these states can be seen in Ref. \cite{Nab99a}.

The total $\beta$-decay rate per unit time per nucleus is finally
given by
\begin{equation}
\lambda^{\beta} =\sum _{ij}P_{i} \lambda _{ij}^{\beta},
\label{betarate}
\end{equation}
where $P_{i}$ is the probability of occupation of parent excited
states and follows the normal Boltzmann distribution. After the
calculation of all partial rates for the transition $i \rightarrow
j$ the summation was carried out over all initial and final states
until satisfactory convergence was achieved in the rate calculation.
The pn-QRPA theory allows a microscopic state-by-state calculation
of both sums present in Eq. ~(\ref{betarate}). This feature of the
pn-QRPA model greatly increases the reliability of the calculated
rates over other models in stellar matter where there exists a
finite probability of occupation of excited states.

Stellar $\beta$-decay rates are sensitive functions of the available
phase space, ($ Q_{\beta} +E_{i} - E_{j}$). It is very much possible
for nuclei with negative $Q_{\beta}$ values to undergo
$\beta$-decays in stellar environment (strictly forbidden under
terrestrial conditions) as the phase space can become positive
depending on the calculated energy eigenvalues of the underlying
theoretical model. The calculated $\beta$-decay rates of
$^{51,52,53,54,55,56}$Ti in stellar environment are presented in
Table 1.
\begin{table}[pt]
\tbl{$\beta$-decay rates of $^{51,52,53,54,55,56}$Ti for selected
densities and temperatures in stellar matter. log$\rho Y_{e}$ has
units of $g/cm^{3}$, where $\rho$ is the baryon density and $Y_{e}$
is the ratio of the lepton number to the baryon number. Temperatures
($T_{9}$) are given in units of $10^{9}$ K. The calculated Fermi
energy is denoted by $E_{f}$ and is given in units of MeV. All
calculated $\beta$-decay rates are tabulated in logarithmic (to base
10) scale in units of $s^{-1}$. In the table, -100 means that the
rate is smaller than 10$^{-100} s^{-1}$.}
{\scriptsize\begin{tabular}{|ccc|cccccc|} $log\rho Y_{e}$ & $T_{9}$
& $E_{f}$& $^{51}$Ti & $^{52}$Ti & $^{53}$Ti & $^{54}$Ti & $^{55}$Ti
& $^{56}$Ti \\ \hline

     1.0  &  0.01  &   0.508  &   -2.698  &   -3.465  &   -2.137  &   -0.466  &   -0.146  &    1.004  \\
     1.0  &  0.10  &   0.453  &   -2.732  &   -3.465  &   -2.137  &   -0.466  &   -0.146  &    1.004  \\
     1.0  &  0.20  &   0.377  &   -2.777  &   -3.465  &   -2.135  &   -0.466  &   -0.146  &    1.004  \\
     1.0  &  0.40  &   0.205  &   -2.812  &   -3.465  &   -2.106  &   -0.466  &   -0.145  &    1.004  \\
     1.0  &  0.70  &   0.008  &   -2.830  &   -3.465  &   -2.049  &   -0.466  &   -0.108  &    1.004  \\
     1.0  &  1.00  &   0.000  &   -2.834  &   -3.465  &   -2.012  &   -0.466  &   -0.009  &    1.004  \\
     1.0  &  1.50  &   0.000  &   -2.807  &   -3.466  &   -1.979  &   -0.466  &    0.169  &    1.004  \\
     1.0  &  2.00  &   0.000  &   -2.739  &   -3.468  &   -1.958  &   -0.466  &    0.292  &    1.004  \\
     1.0  &  3.00  &   0.000  &   -2.557  &   -3.482  &   -1.798  &   -0.468  &    0.437  &    1.003  \\
     1.0  &  5.00  &   0.000  &   -2.074  &   -3.220  &   -0.913  &   -0.471  &    0.594  &    1.002  \\
     1.0  & 10.00  &   0.000  &   -1.031  &   -0.946  &    0.126  &    0.139  &    0.814  &    1.269  \\
     1.0  & 30.00  &   0.000  &    0.221  &    0.699  &    1.172  &    1.424  &    1.616  &    2.138  \\
     4.0  &  0.01  &   0.523  &   -2.699  &   -3.468  &   -2.139  &   -0.466  &   -0.146  &    1.004  \\
     4.0  &  0.10  &   0.516  &   -2.733  &   -3.467  &   -2.138  &   -0.466  &   -0.146  &    1.004  \\
     4.0  &  0.20  &   0.498  &   -2.778  &   -3.466  &   -2.136  &   -0.466  &   -0.146  &    1.004  \\
     4.0  &  0.40  &   0.444  &   -2.813  &   -3.466  &   -2.107  &   -0.466  &   -0.145  &    1.004  \\
     4.0  &  0.70  &   0.337  &   -2.830  &   -3.466  &   -2.050  &   -0.466  &   -0.108  &    1.004  \\
     4.0  &  1.00  &   0.209  &   -2.834  &   -3.466  &   -2.013  &   -0.466  &   -0.009  &    1.004  \\
     4.0  &  1.50  &   0.047  &   -2.807  &   -3.466  &   -1.979  &   -0.466  &    0.169  &    1.004  \\
     4.0  &  2.00  &   0.014  &   -2.739  &   -3.468  &   -1.958  &   -0.466  &    0.292  &    1.004  \\
     4.0  &  3.00  &   0.004  &   -2.557  &   -3.482  &   -1.798  &   -0.469  &    0.437  &    1.003  \\
     4.0  &  5.00  &   0.001  &   -2.074  &   -3.220  &   -0.913  &   -0.471  &    0.594  &    1.002  \\
     4.0  & 10.00  &   0.000  &   -1.031  &   -0.946  &    0.126  &    0.139  &    0.814  &    1.269  \\
     4.0  & 30.00  &   0.000  &    0.221  &    0.699  &    1.172  &    1.424  &    1.616  &    2.138  \\
     7.0  &  0.01  &   1.223  &   -2.918  &   -3.786  &   -2.285  &   -0.543  &   -0.192  &    0.978  \\
     7.0  &  0.10  &   1.222  &   -2.946  &   -3.786  &   -2.285  &   -0.543  &   -0.192  &    0.979  \\
     7.0  &  0.20  &   1.222  &   -2.986  &   -3.785  &   -2.283  &   -0.543  &   -0.192  &    0.979  \\
     7.0  &  0.40  &   1.219  &   -3.015  &   -3.783  &   -2.242  &   -0.543  &   -0.191  &    0.979  \\
     7.0  &  0.70  &   1.212  &   -3.027  &   -3.777  &   -2.165  &   -0.542  &   -0.151  &    0.979  \\
     7.0  &  1.00  &   1.200  &   -3.024  &   -3.768  &   -2.116  &   -0.541  &   -0.044  &    0.979  \\
     7.0  &  1.50  &   1.173  &   -2.980  &   -3.749  &   -2.069  &   -0.538  &    0.143  &    0.980  \\
     7.0  &  2.00  &   1.133  &   -2.886  &   -3.726  &   -2.037  &   -0.535  &    0.272  &    0.981  \\
     7.0  &  3.00  &   1.021  &   -2.658  &   -3.681  &   -1.845  &   -0.528  &    0.422  &    0.983  \\
     7.0  &  5.00  &   0.698  &   -2.109  &   -3.275  &   -0.924  &   -0.507  &    0.585  &    0.989  \\
     7.0  & 10.00  &   0.196  &   -1.037  &   -0.950  &    0.122  &    0.134  &    0.811  &    1.266  \\
     7.0  & 30.00  &   0.021  &    0.221  &    0.699  &    1.171  &    1.423  &    1.615  &    2.138  \\
    10.0  &  0.01  &  11.118  & -100  & -100  & -100  & -100  & -100  & -100  \\
    10.0  &  0.10  &  11.118  & -100  & -100  & -100  & -100  & -100  & -100  \\
    10.0  &  0.20  &  11.118  & -100  & -100  & -100  & -100  &  -87.308  &  -93.598  \\
    10.0  &  0.40  &  11.118  & -100  & -100  &  -74.497  &  -85.977  &  -45.263  &  -48.401  \\
    10.0  &  0.70  &  11.117  &  -62.404  &  -65.400  &  -43.640  &  -50.168  &  -26.906  &  -28.699  \\
    10.0  &  1.00  &  11.116  &  -44.379  &  -46.217  &  -31.133  &  -35.642  &  -19.400  &  -20.626  \\
    10.0  &  1.50  &  11.113  &  -30.186  &  -31.101  &  -21.236  &  -24.133  &  -13.413  &  -14.125  \\
    10.0  &  2.00  &  11.110  &  -22.978  &  -23.418  &  -16.171  &  -18.229  &  -10.322  &  -10.715  \\
    10.0  &  3.00  &  11.099  &  -15.627  &  -15.575  &  -10.948  &  -12.114  &   -7.096  &   -7.084  \\
    10.0  &  5.00  &  11.063  &   -9.553  &   -9.075  &   -6.539  &   -6.902  &   -4.308  &   -3.841  \\
    10.0  & 10.00  &  10.898  &   -4.718  &   -3.878  &   -2.894  &   -2.530  &   -1.871  &   -0.907  \\
    10.0  & 30.00  &   9.163  &   -0.701  &   -0.090  &    0.362  &    0.693  &    0.866  &    1.471  \\
    11.0  &  0.01  &  23.934  & -100  & -100  & -100  & -100  & -100  & -100  \\
    11.0  &  0.10  &  23.934  & -100  & -100  & -100  & -100  & -100  & -100  \\
    11.0  &  0.20  &  23.934  & -100  & -100  & -100  & -100  & -100  & -100  \\
    11.0  &  0.40  &  23.934  & -100  & -100  & -100  & -100  & -100  & -100  \\
    11.0  &  0.70  &  23.934  & -100  & -100  & -100  & -100  & -100  & -100  \\
    11.0  &  1.00  &  23.933  & -100  & -100  &  -95.730  & -100  &  -83.987  &  -85.027  \\
    11.0  &  1.50  &  23.932  &  -73.255  &  -74.170  &  -64.305  &  -67.176  &  -56.474  &  -57.036  \\
    11.0  &  2.00  &  23.930  &  -55.285  &  -55.725  &  -48.478  &  -50.515  &  -42.622  &  -42.893  \\
    11.0  &  3.00  &  23.925  &  -37.175  &  -37.122  &  -32.496  &  -33.647  &  -28.639  &  -28.536  \\
    11.0  &  5.00  &  23.908  &  -22.501  &  -22.023  &  -19.487  &  -19.840  &  -17.252  &  -16.709  \\
    11.0  & 10.00  &  23.832  &  -11.236  &  -10.393  &   -9.411  &   -9.033  &   -8.380  &   -7.344  \\
    11.0  & 30.00  &  23.016  &   -2.980  &   -2.334  &   -1.889  &   -1.527  &   -1.361  &   -0.703  \\
\end{tabular}}
\end{table}
The calculated rates are tabulated on an abbreviated density scale.
The first column gives log($\rho Y_{e}$) in units of $g cm^{-3}$,
where $\rho$ is the baryon density and $Y_{e}$ is the ratio of the
electron number to the baryon number. Stellar temperatures ($T_{9}$)
are stated in $10^{9} K$. The third column shows the calculated
values of the Fermi energy in units of $MeV$. The last six columns
give the calculated $\beta$-decay rates of selected titanium
isotopes in units of $s^{-1}$. The calculated $\beta$-decay rates
are tabulated in logarithmic (to base 10) scale. In the table, -100
means that the rate is smaller than 10$^{-100} s^{-1}$. It can be
seen from the table that at low stellar densities and temperatures,
$^{56}$Ti has the strongest $\beta$-decay rate whereas $^{52}$Ti the
weakest. As the core stiffens from density $\rho Y_{e} [gcm^{-3}]
=10$ to $10^{4}$ the $\beta$-decay rates does not change appreciably
for a particular core temperature. As the core stiffens further the
$\beta$-decay rates start decreasing by orders of magnitude because
of the considerable reduction in available phase space. The
$\beta$-decay rates increase monotonically with increasing
temperature. Positron capture rates act in the same direction as
$\beta$-decay rates and at times tend to compete with the later. The
positron capture rates on these titanium isotopes were also
calculated. Table 2 shows the ratio of the calculated positron
capture to $\beta$-decay rates for isotopes of titanium at selected
temperature and density points.
\begin{table}
\tbl{Ratio of calculated positron capture rates to $\beta$-decay
rates for $^{51,52,53,54,55,56}$Ti for selected densities and
temperatures in stellar matter. log$\rho Y_{e}$ has units of
$g/cm^{3}$, where $\rho$ is the baryon density and $Y_{e}$ is the
ratio of the lepton number to the baryon number. Temperatures
($T_{9}$) are given in units of $10^{9}$ K. The calculated Fermi
energy is denoted by $E_{f}$ and is given in units of MeV. Cases
with no reported ratio imply that either (or both) of the calculated
rate(s) is (are) smaller than 10$^{-100} s^{-1}$.}
{\scriptsize\begin{tabular}{|ccc|cccccc|} $log\rho Y_{e}$ & $T_{9}$
& $E_{f}$& $^{51}$Ti & $^{52}$Ti & $^{53}$Ti & $^{54}$Ti & $^{55}$Ti
& $^{56}$Ti \\ \hline

     1.0  &  0.10  &   0.453  &   1.20E-53  &   5.42E-52  &   3.18E-53  &   5.82E-54  &   2.89E-54  &    1.04E-54  \\
     1.0  &  1.00  &   0.000  &   2.70E-05  &   1.78E-03  &   7.21E-05  &   1.50E-05  &   5.52E-06  &    2.51E-06  \\
     1.0  &  3.00  &   0.000  &   1.46E-02  &   2.32E+00  &   5.20E-02  &   1.23E-02  &   1.79E-03  &    1.75E-03  \\
     1.0  & 10.00  &   0.000  &   1.25E+00  &   2.49E+00  &   4.20E-01  &   7.85E-01  &   1.94E-01  &    1.69E-01  \\
     1.0  & 30.00  &   0.000  &   2.25E+02  &   5.55E+01  &   6.01E+01  &   3.07E+01  &   3.02E+01  &    1.15E+01  \\
     4.0  &  0.10  &   0.516  &   8.04E-57  &   3.64E-55  &   2.13E-56  &   3.90E-57  &   1.93E-57  &    6.98E-58  \\
     4.0  &  1.00  &   0.209  &   2.41E-06  &   1.59E-04  &   6.44E-06  &   1.34E-06  &   4.92E-07  &    2.24E-07  \\
     4.0  &  3.00  &   0.004  &   1.44E-02  &   2.29E+00  &   5.13E-02  &   1.21E-02  &   1.76E-03  &    1.73E-03  \\
     4.0  & 10.00  &   0.000  &   1.25E+00  &   2.49E+00  &   4.20E-01  &   7.85E-01  &   1.94E-01  &    1.69E-01  \\
     4.0  & 30.00  &   0.000  &   2.25E+02  &   5.55E+01  &   6.01E+01  &   3.08E+01  &   3.02E+01  &    1.15E+01  \\
     7.0  &  0.10  &   1.222  &   3.37E-92  &   1.95E-90  &   7.67E-92  &   1.19E-92  &   5.51E-93  &    1.90E-93  \\
     7.0  &  1.00  &   1.200  &   3.75E-11  &   3.21E-09  &   8.20E-11  &   1.60E-11  &   5.37E-12  &    2.38E-12  \\
     7.0  &  3.00  &   1.021  &   3.62E-04  &   7.18E-02  &   1.13E-03  &   2.77E-04  &   3.64E-05  &    3.61E-05  \\
     7.0  & 10.00  &   0.196  &   1.01E+00  &   2.02E+00  &   3.40E-01  &   6.37E-01  &   1.57E-01  &    1.37E-01  \\
     7.0  & 30.00  &   0.021  &   2.24E+02  &   5.51E+01  &   5.98E+01  &   3.05E+01  &   3.01E+01  &    1.14E+01  \\
    11.0  &  0.10  &  23.934  &      -      &      -      &      -      &      -      &     -       &       -      \\
    11.0  &  1.00  &  23.933  &      -      &      -      &      -      &      -      &     -       &       -      \\
    11.0  &  3.00  &  23.925  &   3.97E-08  &   6.61E-07  &   1.69E-11  &   1.21E-09  &   1.40E-14  &   3.97E-14   \\
    11.0  & 10.00  &  23.832  &   1.99E-02  &   7.00E-03  &   1.46E-03  &   1.18E-03  &   3.10E-04  &   7.13E-05   \\
    11.0  & 30.00  &  23.016  &   4.97E+01  &   8.36E+00  &   9.66E+00  &   3.85E+00  &   4.02E+00  &   1.13E+01   \\
\end{tabular}}
\end{table}
It can be seen from this table that only at high stellar
temperatures, T$_{9} [K] \sim 30$,  does the positron capture rate
compete with the $\beta$-decay rates. For all other physical
conditions the positron capture rates are smaller by many orders of
magnitude and hence can be safely neglected as compared to the
$\beta$-decay rates during the presupernova evolution of massive
stars where the core temperature is considerably less. The complete
electronic version (ASCII files) of these rates may be requested
from the corresponding author.

\begin{figure}[th]
\centerline{\psfig{file=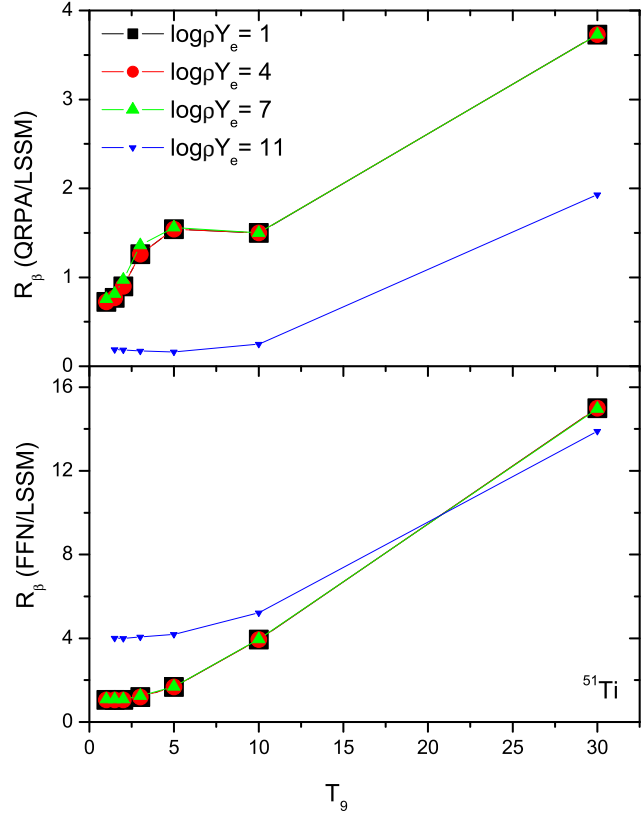, width=10cm}} \vspace*{8pt}
\caption{(Color online) Ratio of reported $\beta$-decay rates to
those calculated using LSSM  (upper panel)
for $^{51}$Ti as function of stellar temperatures and densities.
T$_{9}$ gives the stellar temperature in units of $10^{9}$ K. In the
legend, $log \rho Y_{e}$ gives the log to base 10 of stellar density
in units of $gcm^{-3}$. The lower panel shows the corresponding ratio of FFN $\beta$-decay rates to
those calculated using LSSM.} \label{fig1}
\end{figure}

The calculation of $\beta$-decay rates was also compared with
previous calculations. For the sake of comparison we considered the
pioneer calculations of FFN \cite{Ful80} and those performed using
the large-scale shell model (LSSM) \cite{Lan00}.  Figure~\ref{fig1}
depicts the comparison of $\beta$-decay rates of $^{51}$Ti with
earlier calculations.
The upper panel displays the ratio of calculated rates to the LSSM
rates, $R_{\beta}(QRPA/LSSM)$, while the lower panel shows the corresponding
comparison between FFN and LSSM calculations, $R_{\beta}(FFN/LSSM)$. All
graphs are drawn at four selected values of stellar densities ($\rho
Y_{e} [gcm^{-3}] =10^{1}, 10^{4}, 10^{7}$ and  $10^{11}$). These
values correspond roughly to low, medium-low, medium-high and high
stellar densities, respectively. The selected values for temperature
on the abscissa are T$_{9} [K] = 1, 1.5, 2, 3, 5, 10$ and $30$. It
can be seen from Figure~\ref{fig1} that the pn-QRPA calculated
$\beta$-decay rates of $^{51}$Ti are generally in good agreement
with LSSM rates (within a factor 5).  The comparison between FFN and LSSM calculations is also good except at high temperatures and densities. FFN did not take into
effect the process of particle emission from excited states and
their parent excitation energies extended well beyond the particle
decay channel. The occupation probability of these high lying excited states becomes finite at high temperatures and densities. FFN rates are consequently bigger by more than one order of magnitude at T$_{9} [K] = 30$.

\begin{figure}[th]
\centerline{\psfig{file=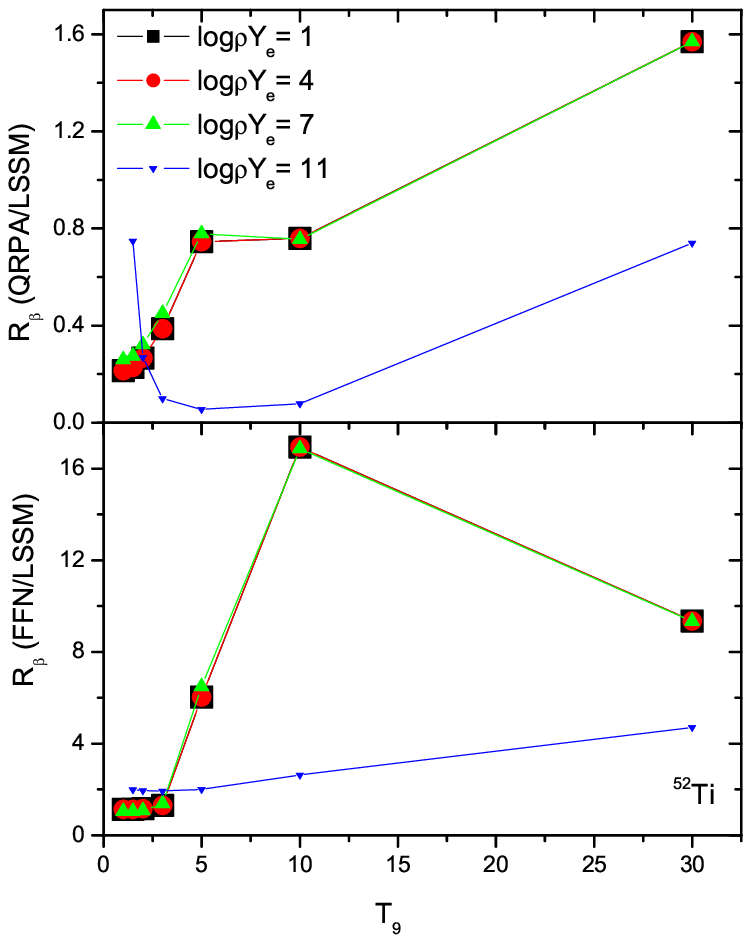, width=10cm}} \vspace*{8pt}
\caption{(Color online) Same as Fig.1. but for $\beta$-decay of
$^{52}$Ti} \label{fig2}
\end{figure}

For the case of $\beta$-decay rates of $^{52}$Ti, the LSSM rates are
slightly bigger (Figure~\ref{fig2}) compared to the reported decay rates.
At low densities the LSSM decay rates are bigger by up to a factor
of five whereas at high densities they are bigger by as much as
factor of twenty. Whereas the individual
discrete transitions between initial and final states matter at low
temperatures and densities, it is the total B(GT) strength that matters
at high temperatures and densities.
Brink's hypothesis is not assumed in this calculation (which was
adopted in LSSM calculation of weak rates). At high temperatures, T$_{9} [K] = 30$, the
agreement is very good hinting towards the fact that the total B(GT) strength calculated in both models match reasonably well.  The lower panel shows that FFN rates are again in reasonable comparison with LSSM rates except at high temperatures for reasons mentioned above.

\begin{figure}[th]
\centerline{\psfig{file=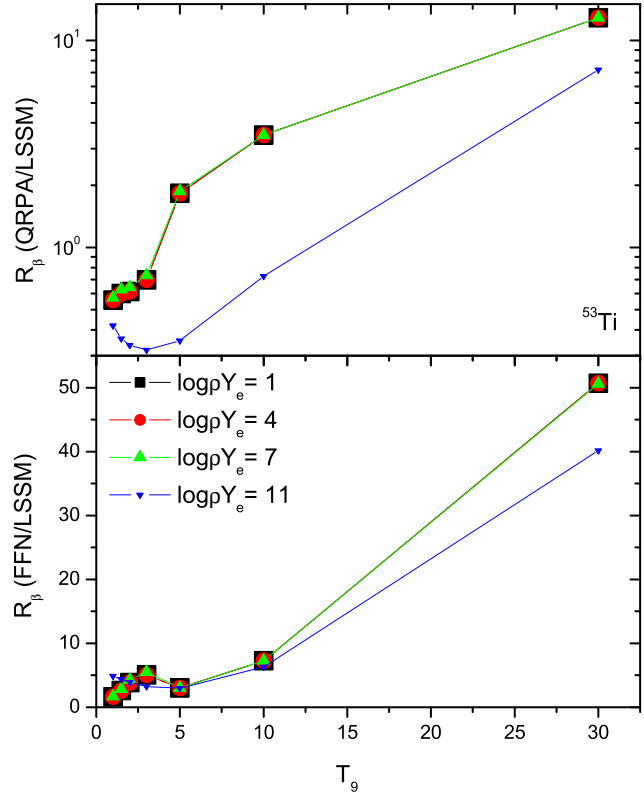, width=10cm}} \vspace*{8pt}
\caption{(Color online) Same as Fig.1. but for $\beta$-decay of
$^{53}$Ti} \label{fig3}
\end{figure}

The agreement with LSSM $\beta$-decay rates for the odd-A nucleus,
$^{53}$Ti, is reasonable (within a factor 5) and depicted in Figure~\ref{fig3}.
However the pn-QRPA rates are bigger by a factor of thirteen at
high temperatures.  Comparison between FFN and LSSM calculations (lower panel) reveals that FFN $\beta$-decay rates are much bigger specially at
high temperatures. Authors in Ref. \cite{Lan00} reported that, for
even-even and odd-A nuclei, FFN systematically placed the back
resonance at much lower energies and concluded
that contribution of the back resonances to the $\beta$-decay rates
for these nuclei decreases. They estimated that LSSM $\beta$-decay
rates as a result were smaller, on the average, by a factor of 20
(40) as compared to the FFN $\beta$-decay rates for even-even
(odd-A) nuclei.

FFN did not calculate the $\beta$-decay rates of $^{54}$Ti and as
such a mutual comparison of the three calculations was not possible
for this even-even isotope of titanium.
\begin{figure}[th]
\centerline{\psfig{file=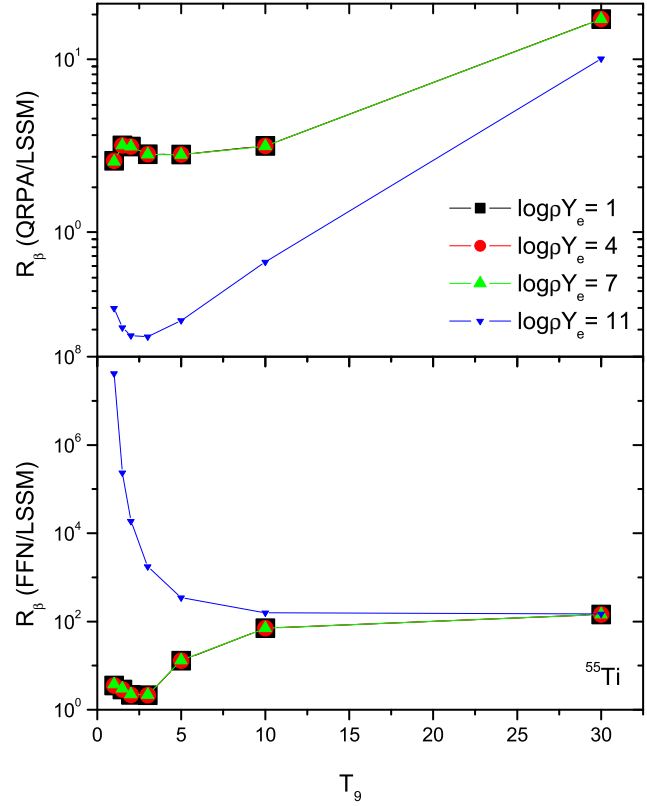, width=10cm}} \vspace*{8pt}
\caption{(Color online) Same as Fig.1. but for $\beta$-decay of
$^{55}$Ti} \label{fig4}
\end{figure}
We note from Figure~\ref{fig4} that our $\beta$-decay rates of
$^{55}$Ti are slightly bigger (factor of 2~-~3) than the LSSM rates
at $\rho Y_{e} [gcm^{-3}] =10^{1}, 10^{4}, 10^{7}$. At high
densities the LSSM rates exceed the pn-QRPA rates as in previous cases. At high stellar
temperatures, the reported rates are more than an order of magnitude
bigger than LSSM $\beta$-decay rates. At higher temperatures excited
state GT strength distributions are required for the calculation of
weak rates (parent excited states have a finite probability of
occupation). The LSSM employed the Brink's hypothesis in
the electron capture direction and back-resonances in the
$\beta$-decay direction to approximate the contributions from
high-lying excited state GT strength distributions.  On the other
hand the pn-QRPA model performs a microscopic calculation of the GT
strength distributions for \textit{all} parent excited states and
provides a fairly reliable estimate of the total stellar rates. The
lower panel of Figure~\ref{fig4} shows a whopping seven orders of
magnitude bigger FFN decay rates at low temperatures and high
densities. However for the same temperature and density domain the
two microscopic calculations (pn-QRPA and LSSM) are in good agreement hinting towards the fact that FFN overestimated their
$\beta$-decay rates by orders of magnitude.  The
centroids of the GT distribution functions determine the
effective energy of the capture and decay reactions. FFN estimated
the GT contributions to the rates by a parametrization based on the
independent particle model and estimated the
GT centroids using zeroth-order ($0\hbar\omega$ ) shell model. Stellar decay rates are fragile functions of the available phase
space, ($ Q_{\beta} +E_{i} - E_{j}$). It is worth mentioning that these $\beta$-decay
rates can change by
orders of magnitude by a mere change of 0.5 MeV, or less in available phase space and are more reflective of the
uncertainties in the calculation of energies. It is to be noted that whereas pn-QRPA and LSSM used the value of 7.34 MeV and 7.81 MeV, respectively, as Q-value for this decay reaction, FFN used a much larger value of 8.64 MeV for the same reaction.

\begin{figure}[th]
\centerline{\psfig{file=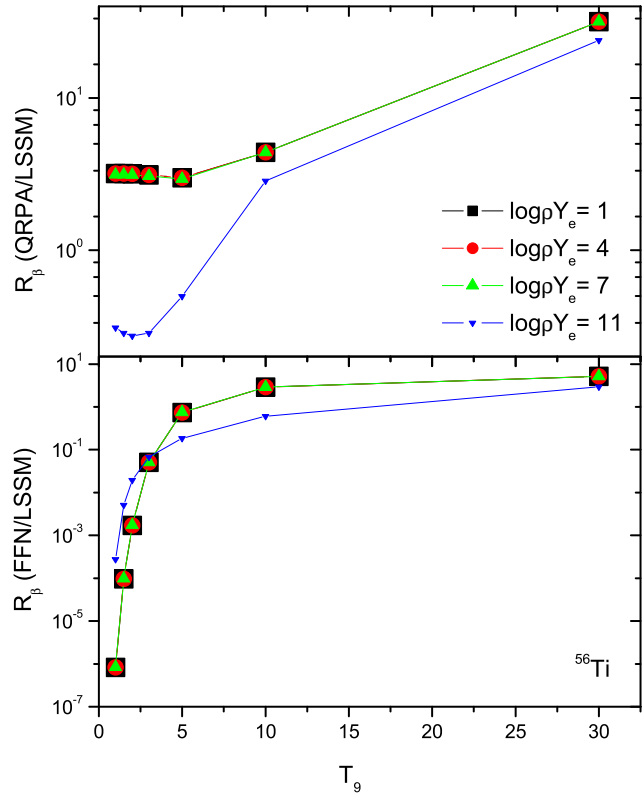, width=10cm}} \vspace*{8pt}
\caption{(Color online) Same as Fig.1. but for $\beta$-decay of
$^{56}$Ti} \label{fig5}
\end{figure}

The comparison of LSSM and pn-QRPA $\beta$-decay rates of $^{56}$Ti
is similar to the previous case of $^{55}$Ti (see
Figure~\ref{fig5}).
At T$_{9} [K] = 30$ the reported decay rates are around a factor of
thirty bigger than LSSM rates. For a change we notice in the lower
panel that the FFN decay rates of $^{56}$Ti are much smaller than the LSSM rates. The
rates are more than six orders of magnitude smaller at
low temperatures and densities. For the same physical conditions the reported
rates are again in far better agreement with the LSSM numbers again
hinting toward some problems in the FFN calculations. It was found
by authors in Ref. \cite{Auf94} that the approach of FFN was not
always reliable in its estimates of the location of GT strength.
Unmeasured matrix elements for allowed transitions were assigned an
average value of $log ft = $5 in FFN calculations. Measurements were performed for the $\beta$-decay of $^{56}$Ti \cite{Doe96} after the FFN calculations. This measurement was properly incorporated in the microscopic calculation of pn-QRPA (this work) and LSSM and resulted in a much bigger  $\beta$-decay of $^{56}$Ti.

\section{Conclusions}

Titanium isotopes are amongst the key iron-regime nuclei that play a
significant role in the late phases of stellar evolution of massive
stars. The  $\beta$-decay rates of these isotopes increase the value
of the lepton-to-baryon fraction ($Y_{e}$) during the late phases of
stellar evolution. The temporal variation of $Y_{e}$ within the core
of a massive star has a pivotal role to play in the stellar
evolution and a fine-tuning of this parameter at various stages of
presupernova evolution is the key to generate an explosion. Six
isotopes of titanium, $^{51,52,53,54,55,56}$Ti, were short-listed as
important $\beta$-decay nuclei as the product of their abundance and
$\beta$-decay rates can cause a substantial change in the time rate
of change of lepton-to-baryon fraction. The pn-QRPA model was used
to calculate the $\beta$-decay rates of these six titanium isotopes
in a microscopic fashion. The pn-QRPA model has two distinct and
important advantages as compared to other models. Firstly it can
handle any arbitrarily heavy system of nucleons since the
calculation is performed in a luxurious model space of up to 7 major
oscillator shells. Further it is the only available model that can
calculate \textit{all} excited state GT strength distributions in a
microscopic fashion which greatly increases its utility in stellar
calculations.

The $\beta$-decay rates were calculated on a detailed
density-temperature grid point. The positron capture rates on
titanium isotopes were also calculated and it was shown that during
the presupernova evolution of massive stars the positron capture
rates can be safely neglected in comparison with the $\beta$-decay
rates. The ASCII files of the calculated rates can be requested from
the corresponding author. The pn-QRPA $\beta$-decay rates were also
compared against previous calculations. Our study validates the
finding of authors of Ref. \cite{Lan00} that FFN systematically
placed the back resonance at much lower energies. Consequently the
FFN overestimated the $\beta$-decay rates for even-even and odd-A
nuclei. The comparison with LSSM calculation is generally fair.
However at high densities the LSSM rates are enhanced whereas at
high temperatures the pn-QRPA $\beta$-decay rates are much bigger.
The study also stresses on the fact that the Brink's hypothesis and
back resonances (employed in previous calculations) are not a good
approximation and it is desirable to microscopically calculate all
excited state GT strength distribution functions for a reliable
estimate of stellar $\beta$-decay rates. Core-collapse simulators
might find it useful to employ the pn-QRPA $\beta$-decay rates as an
alternate (microscopic and reliable) nuclear physics input in their
codes.

\section*{Acknowledgements}
The authors wish to acknowledge the support of research grant
provided by the Higher Education Commission  through the HEC Project
No. 20-1171. JUN also wishes to acknowledge the support of research
grant provided by the Higher Education Commission Pakistan,  through
the HEC Project No. 20-1283.

\end{document}